\documentclass[aps,prl,floatfix,twocolumn,footinbib]{revtex4-2}
\usepackage[utf8]{inputenc}
\usepackage{graphicx}
\usepackage{amsfonts}
\usepackage{amssymb}
\usepackage{amsmath}
\usepackage{bm} 

\newcommand{\eq}[1]{(Eq.\ref{#1})}

\newcommand{\ve}{\bm}
\newcommand{\be}{\begin{equation}}
\newcommand{\ee}{\end{equation}}

\usepackage{color}

\parskip=\medskipamount

\begin{document}

\title{Topological and non-topological mechanisms of loops formation in chromosomes: effects on the contact probability} 

\author{Kirill Polovnikov$^{1,2}$, Bogdan Slavov$^{2}$}

\affiliation{$^1$ Institut Curie, PSL Research University, Sorbonne Université, CNRS UMR3664, Paris, France \\ $^2$ Skolkovo Institute of Science and Technology, 121205 Moscow, Russia }

\begin{abstract}
Chromosomes are crumpled polymer chains further folded into a sequence of stochastic loops via loop extrusion. While extrusion has been verified experimentally, the particular means by which the extruding complexes bind DNA polymer remains controversial. Here we analyze the behaviour of the contact probability function for a crumpled polymer with loops for the two possible modes of cohesin binding, topological and non-topological mechanisms. As we show, in the non-topological model the chain with loops resembles a comb-like polymer that can be solved analytically using the quenched disorder approach. In contrast, in the topological binding case the loop constraints are statistically coupled due to long-range correlations present in a non-ideal chain, which can be described by the perturbation theory in the limit of small loop densities. As we show, the quantitative effect of loops on a crumpled chain in the case of topological binding should be stronger, which is translated into a larger amplitude of the log-derivative of the contact probability. Our results highlight a physically different organization of a crumpled chain with loops by the two mechanisms of loops formation.

\end{abstract}

\maketitle

\section{Introduction}

Chromosomal DNA polymers are organized into loops which serve both structural and gene regulatory roles inside the cell nucleus \cite{mirny_solovei,banigan20}. Folding into chromatin loops provides an effective means of compaction of 10cm-long human chromosomes inside a micron-sized nucleus. In particular, it was proposed that upon the entry to mitosis a fluffy interphase chromosome is arranged into a dense array of loops via the mechanism of lengthwise compaction \cite{goloborodko2016}, which results in the bottlebrush structure long observed in mitosis \cite{paulson77,gibcus18}.

Interphase chromosomes are also organized into loops, however, of significantly lower linear density compared to the mitotic chains \cite{mirny_solovei,fudenberg2017,polovnikov22}. The loops are formed by SMC proteins (e.g, cohesin, condensin), and their extruding ability has been independently demonstrated in single-molecule imaging by two groups \cite{terakawa17,ganji}. In human cells cohesin provides the main source of short-scales loops on chromosomes. However, the mechanism of cohesin binding to chromosomes is not known. It has been proposed that cohesin can topologically embrace two strands of DNA, forming a loop of progressively increasing size in time \cite{pradhan22,banigan20}. In this topological binding model cohesin complex resembles a physical ring that can entrap two DNA strands without chemically binding them. Alternatively, in the non-topological model, cohesin operates as a cross-linker: in the base of the loop it creates a bond between two sites of the chain which is moving in the course of extrusion \cite{pradhan22,banigan20}. Which of the models is realized in the cell is still under debate. As we show here, such a subtle difference in the microscopic structure of cohesin complex can affect chromosome organization at all scales.

A crucial difficulty towards the analytical description of a chromosome as a chain with loops is non-ideal statistics of chromosomal polymer. The physical state of a chromosome can be characterized by the scaling of of the average contact probability function $P(s)$ with the contour distance $s = |i-j|$ (an analogue of the return probability of a random walk to the origin). For chromosomes this function can be extracted from Hi-C experiments \cite{lieberman}. As various data suggest, at large scales, in the range of $s=1-5$ Mb, the scaling of $P(s) \sim s^{-1}$, is different from $\sim s^{-3/2}$ as expected for a three-dimensional random walk, indicating that chromosomes are folded into non-ideal states \cite{lieberman,mirny11,halverson14}. This $\approx -1$ exponent of the contact probability corresponds to the fractal dimension $d_f \approx 3$, which is an asymptotic property of topologically crumpled chains. At shorter scales $s < 1$Mb the scaling of $P(s)$ has a characteristic shoulder, which can be recapitulated by the simplest model of a fractal chain with cohesin-mediated loops breaking the scale invariance \cite{polovnikov22}. Importantly, upon elimination of cohesin loops in experiments \cite{hsieh2021,rao17}, the $\sim s^{-1}$ scaling stretches to shorter distances up to $s \approx 40$kb, suggesting that short-scale cohesin loops are formed on top of a crumpled polymer with strong (power-law) correlations along the genomic contour \cite{polovnikov22}.

Here we demonstrate that due to the intrinsic long-range correlations in folding of crumpled chromosomes the two models of cohesin binding produce drastically different relaxation behaviour of the chain with loops. In the non-topological model a chromosome resembles a comb-like polymer with a backbone and loopy side chains. Conformation properties of the chain at large scales are controlled by the backbone (the shortest path) in this model. In contrast, in the topological binding model the backbone is not formed and the polymer remains linear, while folded into loops. Importantly, in this case the chain length is not shortened by the addition of the loops. Furthermore, all the loops are statistically coupled with each other in the topological model, being the segments of a single crumpled polymer. 
Note that this physical difference between the models vanishes in the case of ideal chain statistics, $d_f=2$, due to absence of long-range correlations in the chain. 

We investigate the effect of the specific binding mechanism on the resulting contact probability function, $P(s)$, for the two models. We treat the loops as frozen disorder on the fractal polymer \cite{polovnikov22} and assume formation of a well-defined sequence of loops and gaps on the chain with exponential distribution of their lengths (no nested or overlapping loops are allowed) fixed on each conformation.  
In order to take into account intrinsic correlations in the crumpled chains, we make use of the Gaussian measure of fractional Brownian paths (fBm) as recently proposed \cite{polovnikovsm18}. We show that while the non-topological model under these assumptions can be computed analytically \cite{polovnikov22}, the topological model is not fully analytically tractable. For the latter case it is possible to construct a perturbation by a single loop (the one-loop approximation), and analytically compute the first order correction to the power-law scaling of $P(s)$. Despite of the lack of the full analytical solution for the topological model, our results show how the two models generate different statistics of contacts across scales.


\section{A fractal polymer folded into random loops}

We exploit a conventional bead-spring model of a flexible polymer chain, generalized for the arbitrary fractal dimension $d_f$, i.e. the average square of the segment size of length $s$ is
\be
r^2_0(s)=b^2 s^{2/d_f},
\label{eq:1}
\ee
where $b$ is a scale of a single bead. In the case of absence of any interactions between the beads other than harmonic polymer bonds, the statistics of chain is ideal with $d_f=2$ \cite{GKh94}. A swollen chain in a good solvent in three dimensions has fractal dimension $d_f \approx 1.7$ \cite{DeGennes1979}, however, excluded volume interactions are likely screened at the relevant scales for chromatin \cite{halverson14}. Topological constraints (in 3D) in the fractal globule model force the chain to fold with the asymptotic fractal dimension $d_f = 3$ \cite{grosberg88,grosberg93}. Without excluded volume $d_f$ can be larger than the dimension of the space: e.g., a phantom randomly-branching polymer has $d_f=4$ \cite{nechaev87}. In general, in order to describe a polymer with any fractal dimension $d_f \ge 2$ an effective quadratic Hamiltonian was proposed in \cite{polovnikovsm18}. It generates Gaussian polymer conformations, which map onto trajectories of a fractional Brownian particle (fBm).

\begin{figure}
 \includegraphics[width=250pt]{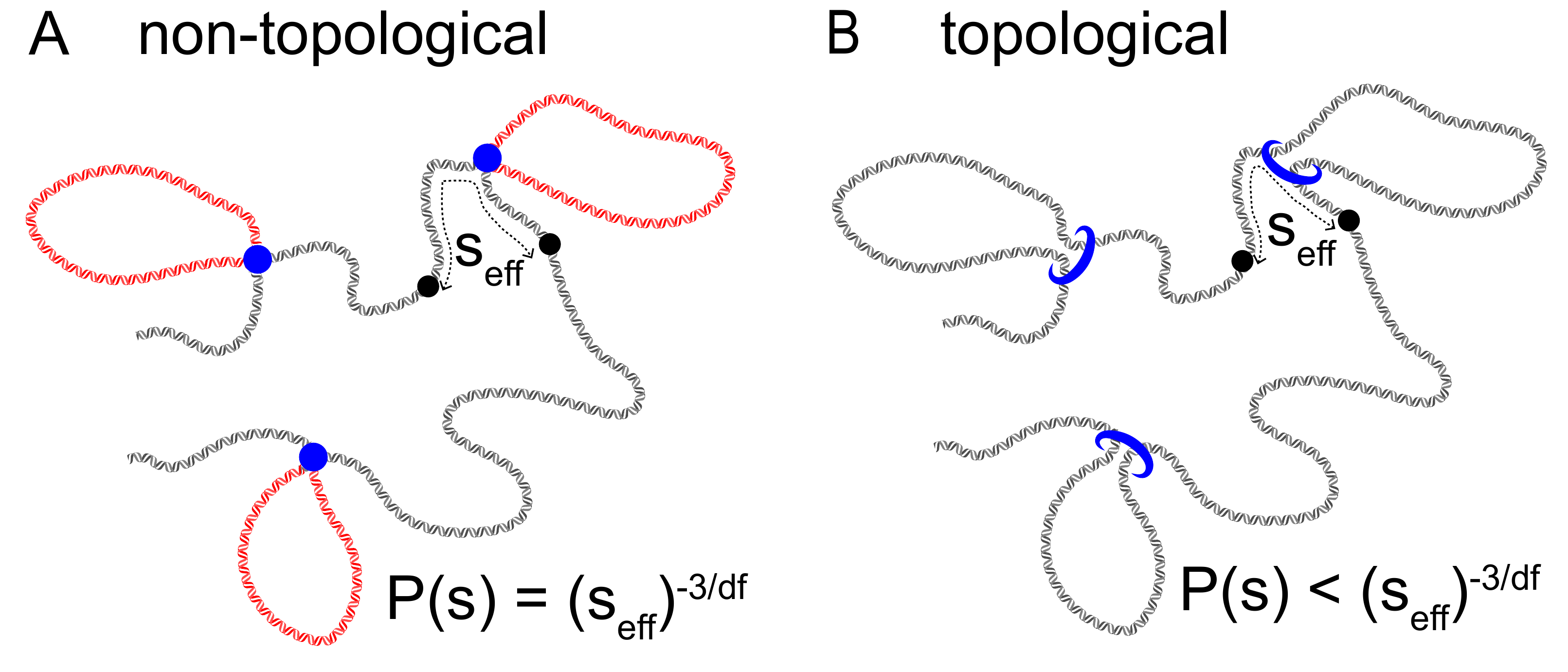}
 \caption{Two possible binding mechanisms of cohesin to DNA. (A). Non-topological model. (B). Topological model.}
  \label{fig1}
 \end{figure}
 
Next we consider an ensemble of fractal chains folded into sequences of loops and gaps, see Fig. 1(b). The loops on each chain in the ensemble are randomly positioned, have a random contour length drawn from exponential distribution with the average $\lambda$ and are separated by exponentially distributed gaps (spacers) with the average length $g$. The loops are fixed for each chain. Similarly to our earlier work \cite{polovnikov22}, we follow the frozen disorder approach, i.e. each chain in the ensemble is \textit{thermally equillibrated} together with its set of the loops. Importantly, the fractal statistics \eq{eq:1} is considered as the inherent property of the chain, owned to specific interactions in the polymer (e.g. large-scale topological constraints), which are not perturbed by addition of the loops. 

In this paper we focus on the difference between the two models of the loop formation, depicted in Fig. 1. In the non-topological binding model cohesin complex forms a covalent bond in the loop base. Therefore, a chain with loops can be partitioned into a fBm backbone (main chain) and a set of fBm bridges (loops). Notably, the main chain and the loops are fractal (generally, non-ideal) chains both having fractal dimension $d_f$ and statistically independent from each other. Relaxation of the polymer at large scales takes place along the shortest path, i.e along the backbone ($s_{eff}$, see Fig. 1A). In contrast, in the case when the ring-shaped binding protein can embrace two strands of chromatin (topological model of binding, Fig. 1B), the chromosome backbone is not formed and stress in the chain propagates along the whole polymer contour. A spacer and its neighbouring loop belong to the same fBm polymer, which implies strong (power-law in the contour distance) tangent correlations between them for $d_f>2$. Thus, these correlations are not destroyed by addition of the loops, which physically determines the difference in the contact statistics of a chain with loops at small and large scales.

\begin{figure}
 \includegraphics[width=200pt]{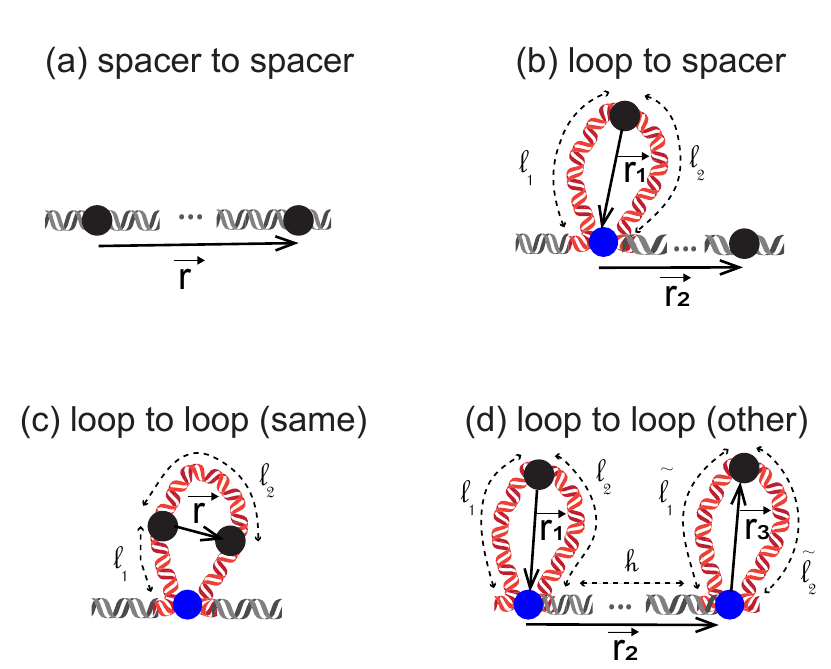}
 \caption{Non-topological model. Diagrams contributing to the contact probability, $P(s)$: (a) spacer-spacer, (b) loop-spacer and (c) loop-loop (same), (d) loop-loop (other).}
  \label{fig2}
 \end{figure}

\subsection{Non-topological model}

Decoupling of correlations between the backbone and the loops in the non-topological model allows for the analytical expression for the $P(s)$, which was derived recently by us in \cite{polovnikov22}. We briefly describe the main steps and results below.

First, one needs to compute the equilibrium contributions of different diagrams, which classify relative positions of the points $i, j$ ($s=|i-j|$) with respect to the loop bases: (a) spacer-to-spacer, (b) loop-to-spacer, (c) loop-loop (same), (d) loop-loop (different). For each diagram one computes the variance of the spatial distance $\sigma^2(i,j)=\langle r^2(i,j) \rangle$ between the points $i, j$ of interest and makes use of the Gaussian relation $P(i, j) \sim \sigma^{-3/2}(i,j)$ for the corresponding equilibrium contact probability. For that the vector $\vec{r}$ is decomposed into the sum of independent vectors (see Fig. 2) ${\vec{r}}_1$ (loop), ${\vec{r}}_2$ (backbone) for the diagram (b); ${\vec{r}}_1$ (loop), ${\vec{r}}_2$ (backbone), ${\vec{r}}_3$ (loop) for the diagram (d). In the diagrams (b)-(d) a loop resembles a fBm bridge of the same dimension $d_f$. The analytical tractability of this model is owned to the statistical independence of the vectors ${\vec{r}}_i, i=1,2,3$, as they correspond to separate branches (side loops and the backbone) in the comb-like polymer. Second, one averages the resulting contact probabilities $P(i, j)$ over all possible pairs of monomers $i, j$ belonging to different diagrams, such that $|i-j|=s$, using the appropriate weights of the diagrams. The exponential distribution of loops and gaps sizes allows to make use of the well-known result for the propagators of the two-state Markov process and properly weigh contributions of different diagrams \cite{Pedler1971}. Finally, the remaining averaging over the distribution of random loops and gaps is performed, which results in the sum of multiple integrals involving Bessel functions \cite{polovnikov22}. 

 Let us see how $P(s)$ for different diagrams look like in this model. As noted above, the variance of the vector connecting any two points at distance $s$ in the chain can be expressed through the variances for a segment free of loops, $\sigma^2_{\text{free}}(s)$ (the backbone) and for a fBm bridge of size $L$, $\sigma^2_{\text{bridge}}(s, L)$. For a free segment along the backbone the result is given by \eq{eq:1}, i.e.
 \be
 \sigma^2_{\text{free}}(s) = b^2 s^{2/d_f},
 \ee
 as the backbone constitutes a fBm trajectory, which is statistically independent of the intervening of loops. The bridge of size $L$ with the fractal dimension $d_f$ can be described as a fBm polymer with the same dimension conditionally constrained by the loop. Making use of the Gaussian measure of the fBm paths, one can derive the following result for the fBm bridge (see \cite{polovnikov22} and derivation of the conditional propagators for the topological model below)
\begin{equation}
  \begin{aligned}
  &\sigma^2_{\text{bridge}}(s, L) = \\
  &= \sigma^2_{\text{free}}(s) \left(1 - \frac{(\sigma^2_{\text{free}}(L)+\sigma^2_{\text{free}}(s)-\sigma^2_{\text{free}}(L-s))^2}{4\sigma^2_{\text{free}}(s)\sigma^2_{\text{free}}(L)}\right),
  \end{aligned}
\end{equation}
 Note, that for the ideal chain $d_f=2$, one obtains the well-known expression for the ideal bridge
\begin{equation}
  \sigma^2_{\text{bridge}}(s, L) = \sigma^2_{\text{free}}(s)\frac{L-s}{L}=\frac{b^2s(L-s)}{s}.
\end{equation}

Now let us consider the diagram (a). 
In the presence of arbitrary number of non-topological loops between two points of interest the contact probability reads 
\begin{equation}
    \label{pa}
    P_{(a)}(s,x)=\frac{1}{(2\pi \sigma_{\text{free}}^2[(1-x)s])^{3/2}}, 
\end{equation}
where $x$ ($0\le x<1 $) denotes the fraction of the subchain length occupied by the loops. In other words, the intervening loops lead to reduction of the effective contour distance between points of interest.

Next, let us consider a subchain of length $s$ with one end belonging to the gap region and another end belonging to the loop. The loop containing one of the two sites of interest is parametrized by the lengths $l_1$ and $l_2$ as shown in Fig 2b.
Clearly, the separation vector $\vec r$ between two sites can be represented as a sum of mutually independent zero mean Gaussian random vectors, $\vec r_1$ and $\vec r_2$. Thus, we have 
\begin{equation}
  \label{pb}
  \begin{split}
    & P_{(b)}(s,l_1,l_2,x)=\\
    &=\frac{1}{(2\pi (\sigma_{\text{bridge}}^2[l_2,l_1+l_2]+\sigma_{\text{free}}^2[(1-x)(s-l_2)])^{3/2}},
  \end{split}
\end{equation}
where $0\le x<1$, $l_1\ge 0$ and $0\le l_2 \le s$. As before, $x$ is the fraction of  contour length along the backbone occupied by the loops. 

Now let us consider a subchain located inside the loop, see Fig. 2c. This situation corresponds to a pure bridge, therefore
\begin{equation}
  \label{pc}
    P_{(c)}(s,l_1,l_2)=
    = \frac{1}{(2\pi \sigma_{\text{bridge}}^2[s,l_1+l_2])^{3/2}},
\end{equation}
with the parameters belonging to the range $l_1\ge 0$ and $l_2\ge s$.

Finally, for the diagram (d) the vector $\vec{r}$ is decomposed into the sum of independent Gaussian vectors $\vec{r}_1$ in the first bridge, $\vec{r}_2$ along the backbone and $\vec{r}_3$ in the second bridge. Thus, the contact probability for this diagram reads
\begin{eqnarray}
  \label{pd}
  P_{(d)}(s,l_1,l_2,h,\tilde{l}_1,\tilde{l}_2,x)=
  \frac{1}{(2\pi \sigma_{d}^{2})^{3/2}},
\end{eqnarray}
with the following variance
\begin{align}
    &\sigma^2_{d}=\sigma_{\text{bridge}}^2[l_2,l_1+l_2]+\sigma_{\text{free}}^2[(1-x)h]+\\
    &+\sigma_{\text{bridge}}^2[\tilde{l}_2,\tilde{l}_1+\tilde{l}_2].
  \label{diagd}
\end{align}

Expressions \eq{pa},\eq{pb},\eq{pc} and \eq{pd} determine the contact probabilities for all four classes of diagrams for any loop density $\lambda/g$ and any fractal dimension $d_f$ of the chain. The final result for the contact probability in the non-topological model is given by the weighted sum
\begin{eqnarray}
  \label{main_Eq}
  P^{\text{non-topo}}(s)=\sum_{i=a,b,c,d} \langle p_{(i)}(s\mid \{ A\}_i)\rangle,
\end{eqnarray}
where $\langle p_{(i)}(s\mid \{ A\}_i) \rangle$ are the particular contributions to the contact probability from each diagram $i$, integrated over the possible values of diagram-specific parameters $\{ A\}_i$ and properly weighted. In particular, the exponential distribution of loops and gaps sizes allows to make use the probabilities for the two-state Markov process for calculation of the weights (see \cite{polovnikov22} for details).

\subsection{Topological model}

\begin{figure}
 \includegraphics[width=250pt]{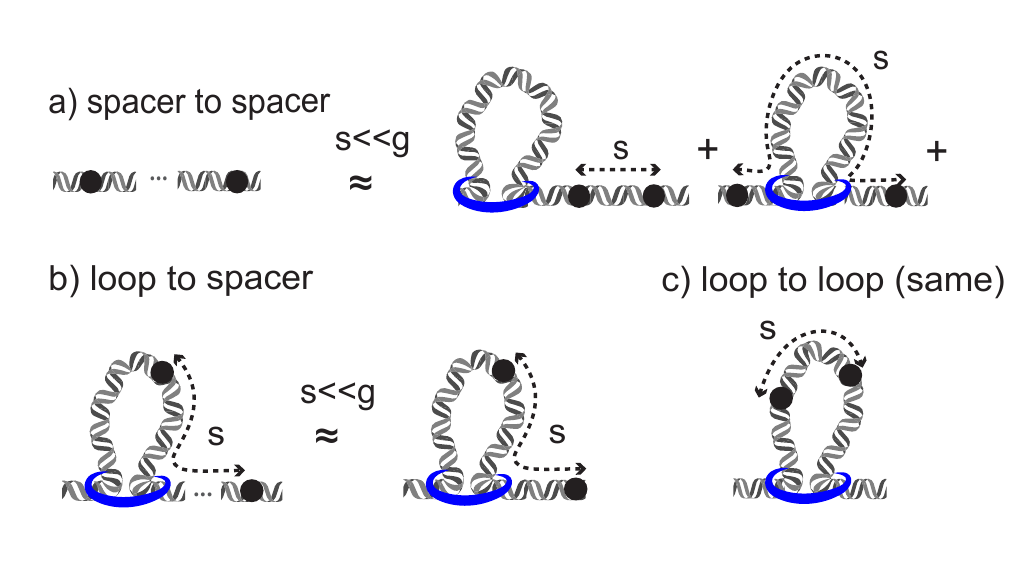}
 \caption{One-loop approximation for the topological model. Diagrams (a) spacer-spacer, (b) loop-spacer and (c) loop-loop (same).}
  \label{fig3}
 \end{figure}

In contrast, in this case all loops and spacers between them are statistically coupled for $d_f \ne 2$, which does not allow to derive the full analytical expression accounting for all the four diagrams. Indeed, in a chain with long-range correlations the conditional probability of contact between any points $i, j$ depends on all the constraints in the chain. In the language of random walks, due to the intrinsic long-range memory in a fBm walk with the Hurst parameter $H>1/2$ ($H=1/d_f$), the conditional return probability at time $t=T$ depends on all the intermediate returns (loops) at previous times $t < T$.

However, treating the loops as a perturbation to the fractal chain scaling, one can obtain a self-consistent linear order correction for $P^{\text{topo}}(s)$ using the Gibbs measure of fBm polymer conformations. For that the non-Markovian propagators of the single loop diagrams (see Fig. 3) should be computed. 

\textit{Propagators of single-loop diagrams}. We have to compute the looping probability of a fBm random walk at time points $i, j>i$ conditioned on the loop at time points $k, n$, such as $k < n$. There are three one-loop diagrams corresponding to this situation that we aim to process (Fig. 3):
\begin{itemize}
\item (a) "spacer-spacer": \\
      (a0) $k<n<i<j$ the points $i,j$ both locate in the gap right next to the loop;\\
      (a1) $i<k<n<j$ the points $i,j$ belong to the consecutive gaps and there is one loop in between them;
\item (b) "loop-spacer": $k<i<n<j$ one point belongs to a loop while the other belongs to the gap next to it;
\item (c) "loop-loop": $k<i<j<n$ both points are inside the loop.
\label{diagrams}
\end{itemize}

\begin{figure*}
\begin{center}
 \includegraphics[width=550pt]{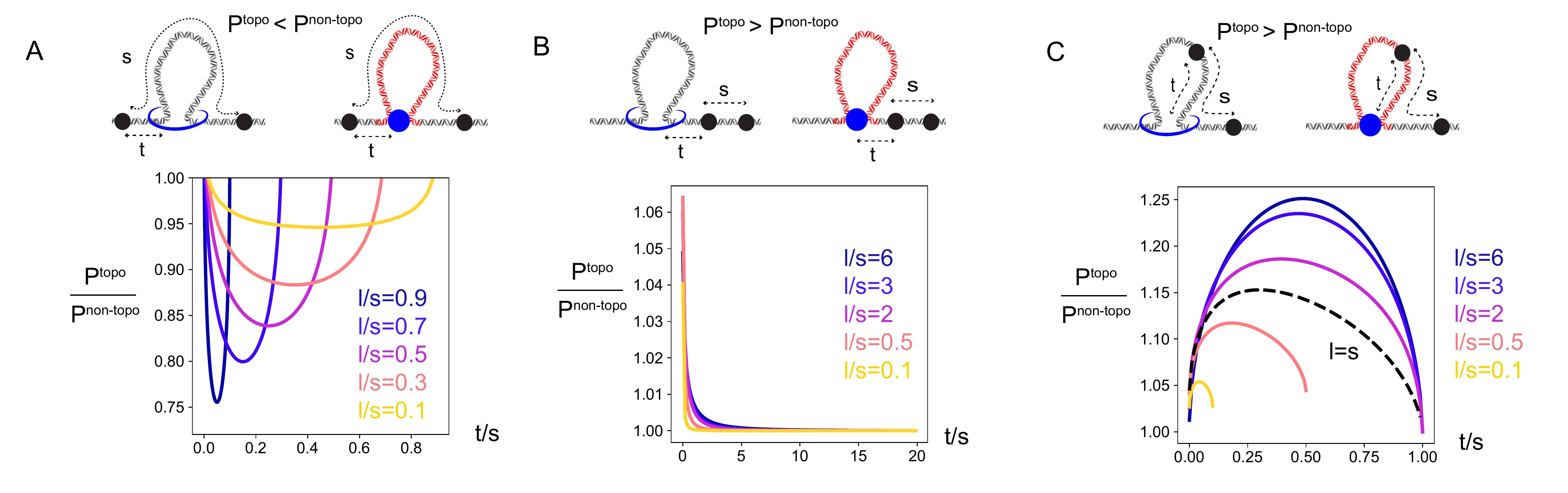}
 \caption{Difference between the contact statistics of the two models for different diagrams and $d_f=3$. On all panels $t$ is the distance between the left point and the loop base. (A). For diagram (a) with one loop in between the points the topological model yields a smaller contact probability than the non-topological. (B). For diagram (a) without loops in between the topological model has a slightly larger $P^{topo}$ than $P^{non-topo}$ for the points located close to the loop base. (C). For diagram (b) the topological model yields a larger contact probability, especially when the loop size is larger than the contour distance between the points, $s<l$.}
  \label{fig4}
  \end{center}
 \end{figure*}
 
The constrained propagators $P(i,j \mid  k,n)$ for the points $i, j$ subject to the loop $k, n$ can be formally written as follows
\be
P(i,j \mid  k,n) = \frac{\int P(\ve X) \delta(\ve x_k - \ve x_n) \delta(\ve x_i - \ve x_j)D\ve X}{\int P(\ve X) \delta(\ve x_k - \ve x_n)D\ve X}
\label{is}
\ee
with the Gibbs distribution $P(\ve X)$ corresponding to the effective Hamiltonian of the fBm polymer. Since all the integrals in \eq{is} are Gaussian, they can computed using the Green function (see our previous works for details \cite{polovnikovpre19,polovnikov22}). The result of the computation reads:
\be
P(i, j \mid  k, n) =  \frac{b^3 (n-k)^{3H}}{(2\pi)^{3/2} 3^{3/2}} \frac{1}{(\det \Sigma)^{3/2}},
\label{sigma}
\ee
where the matrix $\Sigma$ is the covariance matrix built on the vectors $x_i-x_j$ and $x_k-x_n$:
\be
\begin{cases}
\medskip \sigma_{11} = b^2 (n-k)^{2H} \\
\medskip \sigma_{22} = b^2 (j-i)^{2H} \\
\medskip \sigma_{12} = \frac{b^2}{2}\left(|i-n|^{2H} + |j-k|^{2H} - |i-k|^{2H} - |j-n|^{2H}\right).
\end{cases}
\ee
The different diagrams depicted in Fig. 3 correspond to different values of the determinant of the covariance matrix $\Sigma$ in \eq{sigma}. Introducing dimensionless variables $x = 1 - l/s$ and $y = (k-i)/s$, where $s=j-i$ and $l=n-k$, one can express the propagator as follows
\be
P(i, j \mid  k, n) = P(s,x,y) = \frac{1}{s^{3H}} \times I(x, y)
\label{uni}
\ee
where the loops-induced correction function $I(x, y)$ is
\be
I(x, y) \propto \frac{1}{\left(1 - \frac{1}{4}(1-x)^{-2H} g^2(x, y)\right)^{3/2}},
\label{gh}
\ee
and the function $g(x,y)$ has the following universal form
\be
g(x, y) = (1-y)^{2H} + |y+1-x|^{2H} - |x-y|^{2H} - |y|^{2H}.
\label{g}
\ee
The absolute values in \eq{g} are to be expanded depending on relative positions of the monomers in each particular diagram. Clearly, the function $g(x,y)$ reflects the additional damping factor to the unconstrained contact probability $s^{-3H}$ in \eq{uni}, which is due to the loop formation. Importantly, it contains information on the non-Markovian properties of the fractal polymer, specific to the topological binding model.

\textit{The weights of the diagrams in the one-loop approximation}. The weights in the one-loop approximation can be derived in a limit $\lambda/g \ll 1, s/g \ll 1$ of the general expressions in
the full theory, assuming exponential distributions of loop and gap lengths. In the most general terms the weights of each diagram can be calculated as a fraction of configurations of two points $i,j$ at the fixed distance $s=|i-j|$ with respect to the coordinates of the monomers in the loop base $k,n$ that fall into the corresponding diagrammatic class, averaged over all possible loop lengths $l=|k-n|$. For the diagram (a1) with one loop this gives the following result
\be
W_{a_1} = \frac{1}{\lambda+g} \int_0^s dl \rho(l) \int_0^{s-l} dt_1
\label{wa1}
\ee
and $t_1=|k-i|=|y|s$ is the contour distance from the left point to the loop base. The diagram (a0) with no loops in between the points $i,j$ we have two symmetric locations of the points with respect to the loop, thus, it should be taken with the factor of $2$
\be
W_{a_0} = \frac{2}{\lambda+g} \int_0^\infty dl \rho(l) \int_0^{\frac{g-s}{2}} dt_1 = \frac{g-s}{\lambda+g} \approx 1 - \frac{\lambda}{g} - \frac{s}{g}
\label{wa0}
\ee
For the diagram (b) we distinguish two cases $l<s$ and $l>s$ and also have two symmetric cases as for the previous diagram
\be
W_{b} = \frac{2}{\lambda+g} \left( \int_0^s dl \rho(l) \int_0^{l} dt_1 + \int_s^\infty dl \rho(l) \int_{l-s}^{l} dt_1 \right)
\label{wb}
\ee
And the diagram (c)
\be
W_{c} = \frac{1}{\lambda+g} \int_s^\infty dl \rho(l) \int_0^{l-s} dt_1
\label{wc}
\ee

It is easy to check that in the $\lambda/g \to 0$ all the above weights indeed sum to $1$. For the exponential distribution of the loop lengths $\rho(l)=\lambda^{-1} \exp(-l/\lambda)$ the weights can be explicitly calculated
\begin{align}
&W_{a_1} = \frac{s-\lambda}{g} + \frac{\lambda}{g} \exp(-s/\lambda) \\
&W_{a_0} = 1 - \frac{\lambda}{g} - \frac{s}{g} \\
&W_b = \frac{2 \lambda}{g}\left(1 - \exp(-s/\lambda) \right) \\
&W_c = \frac{\lambda}{g} \exp(-s/\lambda)
\end{align}

\textit{Expanding the propagator for each single-loop diagram}. All diagrams can be processed in a similar way, so that the respective weights are effectively encoded in the limits of integration over universal variables $x$ and $y$. For example, for the spacer-spacer diagram (a1) with one loop between points $i, j$ the natural coordinates are the loop size $l$ and the distance from the left point of interest to the left point in the loop base, $t_1 = k-i$
\begin{align}
&P_{a_1} (s) = \frac{1}{s^{3H}}\frac{1}{\lambda+g} \int_0^{s} dl \rho(l) \int_{0}^{s-l} dt_1 I(x(l,s), y(t_1, s))\\
&\approx \frac{1}{s^{3H}} \frac{s}{g} \int_0^1 dx \; \rho(s(1-x)) \int_0^x dy\; I(x,y)
\end{align}
Note that the spacer-spacer diagram (a) without any loops between the points $i,j$ ("free propagator") is trivial only for the Markovian case $H=1/2$. For $H<1/2$ formation of a spatial contact between $i,j$ depends on the fact of loop formation between $k$ and $n$ nearby. Introducing an auxiliary variable $t_1=i-n$ for the distance from the loop to the left point $i$ we can express the propagator as
\be
\begin{split}
&P_{a_0} (s) = \frac{2}{(\lambda+g)s^{3H}} \int_{-\infty}^{1} s dx \; \rho(s(1-x)) \times\\ &\int_{0}^{\frac{g-s}{2}} dt_1 I(x,y(t_1))
\end{split}
\ee
where the upper bound in the second integral takes care of the fact that the main contribution comes from the nearest loop at the distance less or equal than $(g-s)/2$ and the factor $2$ stands for the two symmetric positions of the loop with respect to the points $i,j$ of interest. Switching to the integration over the universal variable $y=-t_1/s+x-1$ we arrive at the following expression
\be
P_{a_0} (s) \approx \frac{2s}{g}\frac{g-\lambda}{g\; s^{3H}} \int_{-\infty}^{1} s dx \; \rho(s(1-x)) \int_{x-1-\frac{g-s}{2s}}^{x-1} dy\; I(x,y)
\ee
which can be further decomposed into Markovian and non-Markovian contributions as follows
\begin{align}
    \label{p0}
    &P_{a_0} (s) \approx \frac{1}{s^{3H}} \left\{ 1 - \frac{\lambda}{g} - \frac{s}{g} + \right.\\
    &\left.+\left(1 - \frac{\lambda}{g}\right)\int_{-\infty}^{1} s dx \; \rho(s(1-x))\; \overline{\Delta I_0(x)} \right\}
\end{align}
where $\overline{\Delta I_0(x)}$ is the typical contribution to the contact probability of the diagram coming from the interdependence of loops and spacers
\be
\overline{\Delta I_0(x)} \approx \frac{2s}{g} \int_{x-1-\frac{g}{2s}}^{x-1} dy \left( I(x,y) - 1\right)
\label{p1}
\ee
Note that the Markovian factor in \eq{p0} is just the weight of the free propagator diagram $\frac{g-s}{g+\lambda} \approx 1 - \lambda/g - s/g$ in the one-loop approximation, see \eq{wa0}. Since the function under the integral in \eq{p1} is integrable, in the limit $s/g \to 0$ the non-Markovian correction is of order of $s/g$. This means we can use $-\infty$ as the lower bound of \eq{p1} and neglect the term $\lambda/g$ in the non-Markovian correction of \eq{p0} to the leading order. Analogously, the propagators for the diagrams (b) and (c) can be written using the corresponding expressions for the conditional probabilities \eq{uni} and the weights \eq{wb} and \eq{wc}.

\textit{The perturbative form of the contact probability}. In the one-loop approximation, the contact probability ${\cal{P}}^{\text{topo}}(s)$ for the topological model is determined by the sum of contributions coming from the diagrams $(a)$, $(b)$ and $(c)$, i.e.
\begin{equation}
\label{contact_probability01}
{\cal{P}}^{\text{topo}}(s) = P_{a_1}(s)+P_{a_0}(s)+P_b(s)+P_c(s),
\end{equation}
so that the resulting expression for the averaged contact probability in the one-loop approximation can be written as follows
\be
{\cal{P}}^{\text{topo}}(s) \propto \left(1 + \frac{\lambda}{g} f(s)\right)\frac{1}{s^{3H}}
\label{res11}
\ee
where the first term reflects the contribution of the free propagator and the function $f(s)$ in the brackets responds for the perturbative contribution of the one-loop diagrams
\begin{align}
    &f(s) = -1+\frac{s}{\lambda}\Biggl(-1+\\
    &2 \int_{-\infty}^{0} s dx \; \rho(s(1-x)) \int_{-\infty}^{x-1} dy \left(I(x,y)-1\right) + \\
    &+\int_0^1 s dx \; \rho(s(1-x)) \left(\int_0^x dy \; I(x, y)
    + 2\int_{x-1}^{0} dy \; I(x, y) \right) \\
    &+\int_{-\infty} ^0 s dx \; \rho(s(1-x)) \left(\int_x^0 dy \; I(x, y) + 2\int_{x-1}^{x} dy \; I(x, y) \right) \Biggr).
    \label{H}
\end{align}

Taking a log derivative of \eq{res11} and keeping only linear in $\lambda/g$ terms, we arrive at the following perturbative expansion
\be
\frac{\partial \log {\cal{P}}^{\text{topo}}(s)}{\partial \log s} = -3H + \frac{\lambda}{g}\Delta(s),
\ee
where $\Delta(s)=s\frac{\partial f(s)}{\partial s}$. It is straightforward to show that $\lim_{s\to0}f(s)=0$ so that the loop-free result gets recovered at small scales (short polymer subchains do not feel the loop constraints). In the opposite limit one obtains $\lim_{s\to\infty} f(s)=3H$.


\begin{figure*}
\begin{center}
 \includegraphics[width=500pt]{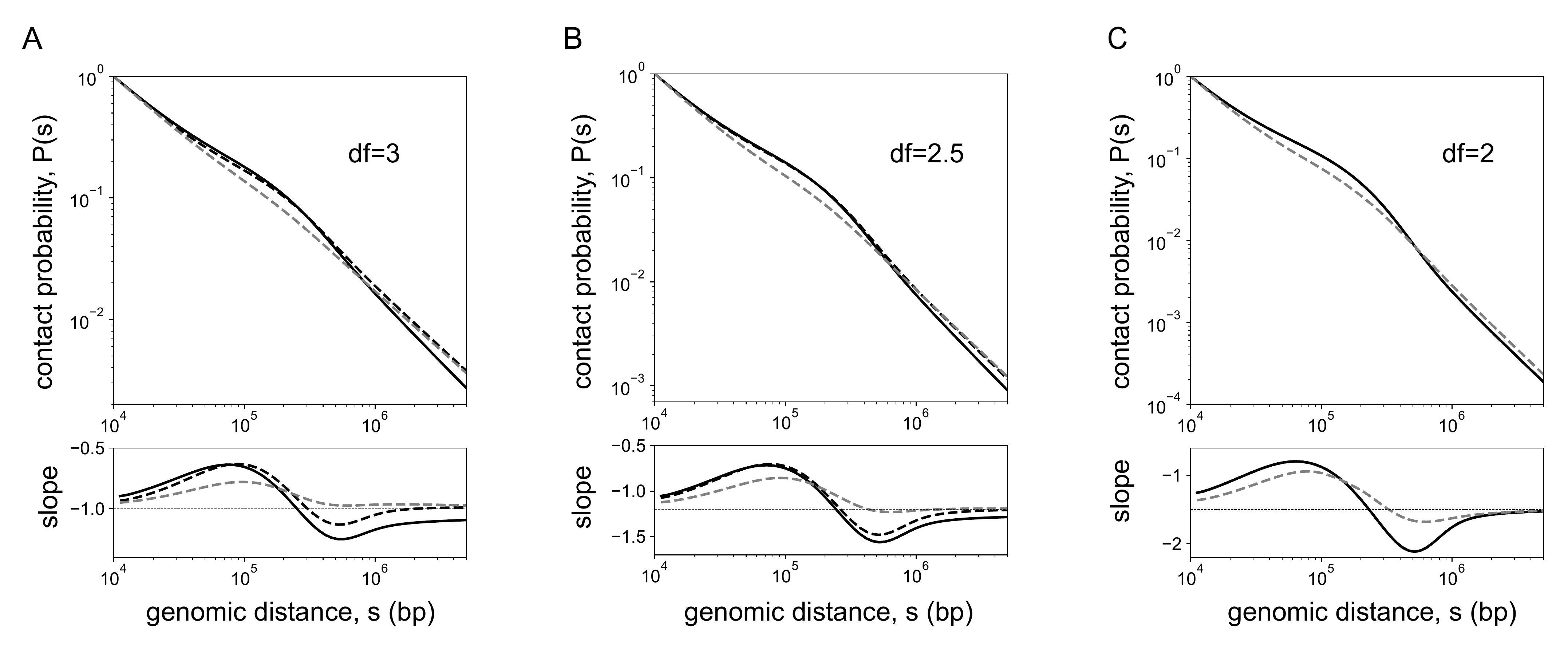}
 \caption{The contact probability and its log-derivative for the non-topological binding model $P^{\text{non-topo}}(s)$ (gray, dashed), for the non-topological in the one-loop approximation ${\cal{P}}^{\text{non-topo}}(s)$ (black, dashed) and for the topological in the one-loop approximation ${\cal{P}}^{\text{topo}}(s)$ (black, solid). Loop density is $\lambda/g=1$. Different fractal dimensions of the chain: (A) $d_f=3$, (B) $d_f=2.5$, (C) $d_f=2$. The difference between the topological and non-topological models vanishes as the chain is becoming more ideal.}
  \label{fig5}
  \end{center}
 \end{figure*}

\subsection{Difference in the shape of $P(s)$ for the two models}

Having the results for the two models, it is important to see how the residual correlations in the topological binding mechanism change the contact probability across scales. To understand it, in what follows we will analyze their impact for the main diagrams (a) and (b). The diagram (c) is not relevant for us here, since it provides equivalent results for the both models. The diagram (d) is not included in the one-loop approximation, which is used to compute the topological model. However, at large contour distances $s$ it can be approximated by the behaviour of the diagram (a). Indeed, contribution of the backbone segment in \eq{diagd} becomes dominant at $s \gg \lambda$. 

The diagram (a) consists of two diagrams: (a0) and (a1). Fig. 4A shows that for the two points located on the backbone and one loop in between them (diagram (a1)) the contact probability is always larger for the non-topological binding of cohesin, $P^{\text{non-topo}}_{a1}>P^{\text{topo}}_{a1}$. Importantly, the difference gets stronger with increase of the loop size $l$ in between. The effect is the most pronounced when the two points are located close to the loop base, $l/s \approx 1$, i.e. when most of the contour distance is occupied by the loop. The source of this difference is clear. From the viewpoint of a random walk, the time elapsed between the points is smaller for the non-topological model as it effectively goes along the backbone ($s_{eff}$). In other words, the effective contour distance between the points in the diagram (a1) is shortened by formation of the loops in between. In the case of topological binding such shortening is absent for $H<1/2$ and the shortest contour distance equals to the original distance $s>s_{eff}$. Therefore, the equilibrium distance between the points is larger in the topological model and the contact probability is smaller. 

However, for $H=1/2$, the difference in $P(s)$ for the diagram (a1) between the models vanishes, despite the shortest path along the polymer in the topological model is larger than in the non-topological. One can demonstrate this as follows. As in the previous section, let us denote by $i, j$ the two points of interest and by $k, n$ the base of the loop. The vector $\vec{r}_{ij}=\vec{r}_{i}-\vec{r}_{j}$ can be decomposed as
\be
\vec{r}_{ij}=\vec{r}_{ik}+\vec{r}_{kn}+\vec{r}_{nj},
\label{rij}
\ee
where $\vec{r}_{kn}=0$ due to the loop. Since in the case of ideal chain all the vectors in \eq{rij} are independent, the distance between the points $i,j$ is controlled by the distances along the backbone
\be
(\vec{r}_{ij})^2 = (\vec{r}_{ik})^2+(\vec{r}_{nj})^2=(k-i)+(j-n),
\ee
i.e. the loop is effectively eaten up. Thus, the shortest distance in the non-topological model becomes the effective contour distance in the topological for $H=1/2$ for the diagram (a1).

Interestingly, for diagrams (b) and (a0) the situation is the opposite (see Fig. 4B,C): in the topological model the contact probability is larger, $P^{\text{topo}}_{a0,b}>P^{\text{non-topo}}_{a0,b}$, in the case of non-ideal chain. Similarly to the diagram (a1), upon increase of the loop size $l$ the effect gets stronger. However, the physics of the difference between the models in this case is not related with the chain shortening, but with residual correlations that are not destroyed by formation of a loop via the topological mechanism. Indeed, for example, for the diagram (b) residual correlations between a point in the loop and a point in the spacer persist in the topological mode of loop formation. Since these correlations are negative (for $H<1/2$), the equilibrium distance between the points is shorter than in the non-topological case, where the points fluctuate independently. Accordingly, for the ideal chain statistics ($H=1/2$) the two points in either model fluctuate independently, and the contact probability is the same. Interestingly, for sufficiently large loops, $l \gg s$, the maximal fold-difference between the models corresponds to the symmetric positions of the points relatively to the loop base, $t/s=1/2$ (see Fig. 4C). However, upon decrease of the loop size $l$ the optimal relative distance to the base drops down to $t/s=1/4$ at $l=s$, meaning that the point in the spacer should be three times further from the loop base than the one in the loop. Such an asymmetry in the configuration is related with different statistical properties of spacer and loop regions, which becomes noticeable when their contour sizes get similar.

Despite the topological model can be solved only in the one-loop approximation (i.e. for $\lambda/g \ll 1$), one can understand the qualitative difference between the models for all $\lambda/g$. Indeed, the conformational statistics of a loopy chain at $s \gg \lambda$ is determined by the distance between the points located on the main chain, i.e. by the diagram (a), see Fig. 2(a). This distance is effectively shortened in the case of non-topological binding model, thus leading to a higher contact probability, similarly to the one-loop diagram (a1), see Fig. 4A.  With increase of the separation $s$ between the points the fraction of the chain within the backbone (along the shortest path) increases to the bulk value $\frac{g}{\lambda+g}$. For the reference loop density $\lambda/g=1$, this corresponds to half of the contour distance effectively shortened. For the case of a single loop diagram (a1), this yields around $\approx 15\%$ for the difference in the absolute values of $P(s)$ at large scales (Fig. 4a). Therefore, with increase of loop density $\lambda/g$ the drop of $P^{\text{topo}}$ compared to $P^{\text{non-topo}}$ at large scales increases, in particular, resulting in a deeper dip of the log-derivative. 

On the other hand, at short scales $s \approx \lambda$ all diagrams depicted in Fig. 4 contribute similarly to $P(s)$. While the diagram (a1) with one loop in between yields a larger value of $P(s)$ in case of non-topological binding, this is not the case for diagrams (b) and (a0). Thus, we expect the difference between the models at short scales to be less pronounced.

This analysis is confirmed by Fig. 5, where for the sake of accuracy we compare the two models in the one-loop approximation ${\cal{P}}^{\text{topo}}(s)$ and ${\cal{P}}^{\text{non-topo}}(s)$, as well as the full theory for the non-topological model, $P^{\text{non-topo}}(s)$. For that we use the expressions \eq{pa}-\eq{pc} for the single-loop diagrams in the non-topological case and insert them as the particular contributions for calculation of ${\cal{P}}^{\text{non-topo}}(s)$. The averaging of the single-loops diagrams, as for the topological model, is conducted using the linearized weights \eq{wa1}-\eq{wc}. First we note that at intermediate loop densities ($\lambda/g=1$) the one-loop approximation produces larger amplitudes of the peak and the dip on the log-derivative, for all fractal dimensions $d_f$. Interestingly, the effect of the topological binding at the loop base is qualitatively similar. Indeed, in the full agreement with our analysis above, if one compares ${\cal{P}}^{\text{topo}}(s)$ and ${\cal{P}}^{\text{non-topo}}(s)$ (black lines in Fig. 5), it is evident that the topological model yields a smaller contact probability at larger scales and, thus, larger amplitude of the dip. The corresponding difference at short scales $s \le \lambda$ is much less pronounced than at large scales. Still one sees that the topological model at short scales produces somewhat larger and slower decaying ${\cal{P}}^{\text{topo}}$, especially for larger values of the fractal dimension. Upon decrease of the fractal dimension from $d_f=3$ to $d_f=2$ the two models in the one-loop approximation give more similar results, consistently with our analysis.

All together, our analysis shows that the main difference between the binding models concerns large scales. With increase of the loop density, the topological model produces smaller $P^{\text{topo}}(s)$ than the non-topological $P^{\text{non-topo}}(s)$ at $s \gg \lambda$, resulting in a deeper dip of the log-derivative. Physically, this difference arises from the effective chain shortening that is established by the loops in the non-topological binding model.

\section{Discussion}

Chromosomes are polymers organized into loops by SMC complexes (e.g. condensins, cohesins) \cite{mirny_solovei}. These loops are transient, as they are produced in a process of loop extrusion of a SMC complex that stochastically binds and unbinds DNA \cite{fudenberg2017}. Therefore, these random loops are not directly visible in the data (Hi-C or microscopy). Here and in our previous paper \cite{polovnikov22} we have suggested an analytically tractable polymer model, in which the presence of random loops, as well as their statistical characteristics, are encoded in the specific shape of the average contact probability curve. At the same time, this $P(s)$ function can be also computed from the experimental Hi-C contact maps \cite{lieberman}, providing a robust framework to characterize the loopy state of a chromosome.

Despite the loop extrusion by cohesin (and, generally, by SMC complexes) has been confirmed in \textit{in vitro} experiments \cite{terakawa17,ganji,kim19,golfier20,davidson19}, the microscopic mechanism of binding and reeling on DNA remains contradictory. As other SMC complexes, cohesin has a ring-shaped structure, but whether this ring is actually needed to topologically embrace two strands of DNA in the course of extrusion is a subject of ongoing experimental and computational research \cite{pradhan22,banigan20}. It has been proposed three possible binding mechanisms \cite{pradhan22}: (i) topological, i.e. the cohesin is bound to one string chemically, while the ring opens and closes when cohesin needs to encircle the second strand of DNA; (ii) pseudo-topological, when both strands are embraced by the ring without chemical binding in the loop base, so that cohesin can be easily pulled away from the chromosome; (iii) non-topological, when the two DNA strands are chemically bound to the ring, which implies the cohesin ring does not play an important role in binding the DNA. In particular, in the mechanisms (i) and (ii) the chromatin strand is reeled into the cohesin ring in the course of extrusion together with the proteins residing on the chain. Clearly, translocation of obstacles larger than the ring size ($\approx 35$nm) into the loop would not be possible. However, a recent single-molecule study has shown that nanoparticles as large as $\approx 200$nm, as well as big protein machines (e.g., RNA polymerase), can be easily translocated into the loop by cohesin \cite{pradhan22}. This experimentally observed ability of cohesin to avoid large obstacles on DNA points to the non-topological binding mechanism. However, other authors have previously noted that the topological binding could explain atypically long residence times of cohesin on chromatin ($\approx 10$ minutes), as compared to other molecules without the ring structure \cite{banigan20}. Furthermore, cohesin was previously shown to topologically entrap two sister chromatids \cite{ivanov05}. Thus, the role of the cohesin ring structure in DNA binding and extrusion of loops is not yet fully established.

In this paper we underscore the importance of a particular microscopic structure of the cohesin loop base on the 3D organization of a chromosome across scales. Standing on the experimental observation that without cohesin loops chromosomes follow the statistics of a crumpled chain with fractal dimension $d_f=3$, and not of an ideal chain with $d_f=2$ \cite{hsieh2021,rao17,polovnikov22}, we demonstrate that the conformational properties of a crumpled chain with loops would be different for the topological and non-topological structure of the loop base. While for the non-topological binding the backbone of a loopy polymer forms with the loops emanating out of it (as in a comb-like polymer), in the topological case the polymer folds into loops, but its organization remains linear. This leads to the two main peculiarities of the two models: (i) effective shortening of the chromosome at large scales in the non-topological model, and (ii) negatively correlated fluctuations of adjacent loops and spacers in the topological model. Notably, for the ideal chain statistics ($d_f=2$) the correlations along the chain are absent, resulting in equivalent organization of the loopy chain in either binding model. However, for any $d_f>2$, due to strong (power-law) negative correlations, the mode of cohesin binding has a quantitative effect on the three-dimensional organization of the chains into loops. 

In order to quantify this effect we used an analytically tractable polymer model \cite{polovnikov22} and compare the behaviour of the contact probability $P(s)$ for topological and non-topological binding models. We based our theoretical argument on the quenched disorder approach and made several assumptions that allowed us to solve the non-topological model analytically for any fractal dimension $d_f$ and any density of loops $\lambda/g$ (see \cite{polovnikov22}). Under the same set of assumptions, the topological model is not analytically solvable, so here we employed the one-loop approximation which treats the loops as a perturbation to the fractal scaling of the chain. We obtained the perturbative form of $P(s)$ in the loop density $\lambda/g$ and analytically computed the linear order correction using the Gaussian measure of fBm trajectories as suggested recently. This result allowed us to quantitatively compare the contact probability between the two models for the single-loop diagrams ((a), (b) and (c)). We found that the binding mechanism in the loop base has indeed the impact on the $P(s)$, which is qualitatively similar to the difference between the one-loop approximation and the full theory within the single binding mechanism.


The revealed quantitative difference for the two binding models allows to rationalize the organization of chromosomes beyond the minimal analytically tractable models. We note the central role of the non-ideal (crumpled) statistics of the chain in the observed difference. Thus, for rare species where the emergent statistics of the chromosomes is closer to the ideal chain (e.g. yeast \cite{wong12}), our analysis implies universal contact probability behaviour independent of the microscopic details imposed by the SMC machinery.

\section{Acknowledgements}

We thank Leonid Mirny, Sergei Nechaev, Mehran Kardar, Hugo Brandao, Vittore Scolari and Sergei Belan for valuable discussions on the subject of the paper. The work is supported by the Russian Science Foundation (Grant No. 21-73-00176).

\section*{References}

\bibliography{bibliography}

\end{document}